\documentclass[prb,final,twocolumn,superscriptaddress]{revtex4}
\usepackage{graphicx}

\begin{document}

\title{Strongly correlated properties of the thermoelectric cobalt oxide Ca$_{3}$Co$_{4}$O$_{9}$}

\author{P. Limelette}
\affiliation{Laboratoire CRISMAT, UMR 6508 CNRS-ENSICAEN et Universit\'e de Caen, 6, Boulevard du Mar\'echal Juin, 14050 CAEN Cedex, France}
\author{V. Hardy}
\affiliation{Laboratoire CRISMAT, UMR 6508 CNRS-ENSICAEN et Universit\'e de Caen, 6, Boulevard du Mar\'echal Juin, 14050 CAEN Cedex, France}
\author{P. Auban-Senzier}
\affiliation{Laboratoire de Physique des Solides (CNRS, U.R.A. 8502), B\^atiment 510, Universit\'e de Paris-Sud, 91405 Orsay, France}
\author{D. J\'erome}
\affiliation{Laboratoire de Physique des Solides (CNRS, U.R.A. 8502), B\^atiment 510, Universit\'e de Paris-Sud, 91405 Orsay, France}
\author{D. Flahaut}
\affiliation{Laboratoire CRISMAT, UMR 6508 CNRS-ENSICAEN et Universit\'e de Caen, 6, Boulevard du Mar\'echal Juin, 14050 CAEN Cedex, France}
\author{S. H\'ebert}
\affiliation{Laboratoire CRISMAT, UMR 6508 CNRS-ENSICAEN et Universit\'e de Caen, 6, Boulevard du Mar\'echal Juin, 14050 CAEN Cedex, France}
\author{R. Fr\'esard}
\affiliation{Laboratoire CRISMAT, UMR 6508 CNRS-ENSICAEN et Universit\'e de Caen, 6, Boulevard du Mar\'echal Juin, 14050 CAEN Cedex, France}
\author{Ch. Simon}
\affiliation{Laboratoire CRISMAT, UMR 6508 CNRS-ENSICAEN et Universit\'e de Caen, 6, Boulevard du Mar\'echal Juin, 14050 CAEN Cedex, France}
\author{J. Noudem}
\affiliation{Laboratoire CRISMAT, UMR 6508 CNRS-ENSICAEN et Universit\'e de Caen, 6, Boulevard du Mar\'echal Juin, 14050 CAEN Cedex, France}
\author{A. Maignan}
\affiliation{Laboratoire CRISMAT, UMR 6508 CNRS-ENSICAEN et Universit\'e de Caen, 6, Boulevard du Mar\'echal Juin, 14050 CAEN Cedex, France}

\begin{abstract}
\vspace{0.3cm}
We have performed both in-plane resistivity, Hall effect and specific heat measurements on the thermoelectric cobalt oxide Ca$_{3}$Co$_{4}$O$_{9}$.
Four distinct transport regimes are found as a function of temperature, corresponding to a low temperature insulating one up to $T_{min}\approx $63 K,
a strongly correlated Fermi liquid up to $T^*\approx $140 K, with $\rho=\rho_0+AT^2$ and $A\approx 3.63$ $10^{-2} \mu \Omega cm/K^{2}$, 
followed by an incoherent metal with $k_Fl\leq 1$ and a high temperature insulator above T$^{**}\approx $510 K . 
Specific heat Sommerfeld coefficient $\gamma = 93$  mJ/(mol.K$^{2}$) confirms a rather large value of the electronic effective mass 
and fulfils the Kadowaki-Woods ratio $A/\gamma^2 \approx  0.45$ 10$^{-5}$ $\mu \Omega cm.K^2/(mJ^2mol^{-2})$.
Resistivity measurements under pressure reveal a decrease of the Fermi liquid transport coefficient A with an increase of $T^*$ as a function of pressure
 while the product $A(T^*)^2/a$ remains constant and of order $h/e^2$. Both thermodynamic and transport properties suggest a strong renormalization of 
the quasiparticles coherence scale of order $T^*$ that seems to govern also thermopower.
\end{abstract}

\maketitle
While the discovery of superconductivity in hydrated  Na$_{0.35}$CoO$_2$  \cite{Takada03} has considerably stimulated 
both experimental \cite{Sugiyama04} and theoretical studies of layered cobalt oxides, the origin of their large thermopower coexisting with metallic properties  \cite{Terasaki97} 
remains currently unclear. Of major interest for saving energy, thermoelectricity appears as a fundamental phenomenon 
which mechanism can involve spin degeneracy \cite{Koshibae00}, charge frustration \cite{Motrunich03} or enhanced effective mass in the vicinity 
of a Mott metal-insulator transition \cite{Palsson98}, both combined with Coulomb repulsion. Thus, it is essential to demonstrate the existence of strong electronic correlations 
in cobalt oxides and to characterize their effects in order to put these scenarios to experimental test.\\
Derived from Na$_{x}$CoO$_2$, the cobaltite Ca$_{3}$Co$_{4}$O$_{9}$ has a misfit structure consisting of single [CoO$_2$] layer of CdI$_2$ type 
stacked with [CoCa$_2$O$_3$] rocksalt type layers, with different in-plane lattice parameter b \cite{Masset00,Lambert01}. As a result, this misfit layered oxide shows 
highly anisotropic properties \cite{Masset00}, including magnetic ones at low temperatures \cite{Sugiyama03}.
Interestingly, while the NaCo$_2$O$_4$ thermopower S increases continuously with temperature up to 100 $\mu V/K$ at 300 K \cite{Terasaki97}, 
the Ca$_{3}$Co$_{4}$O$_{9}$ Seebeck coefficient becomes weakly temperature dependent from 150 K 
and reaches 130 $\mu V/K$ at 300 K \cite{Masset00}.

We report in this letter on a detailed experimental study of both transport and thermodynamic properties of the misfit cobalt oxide Ca$_{3}$Co$_{4}$O$_{9}$ 
 to discuss whether or not strongly correlated features dominate and could explain a high thermopower. 
Temperature dependence of the in-plane single crystal resistivity  from 2 K to nearly 600 K has led us to identify  transport crossovers, 
including a crossover from a Fermi liquid, with $\rho=\rho_0+AT^2$, to an incoherent metal above $T^{*}$. 
This crossover has been analysed from transport experiments under pressure within the range of 1 bar to 12 kbar below 300 K.
By analysing  pressure dependences of Fermi liquid parameters, strong electronic correlations are confirmed.

\begin{figure}[htbp]
\centerline{\includegraphics[width=0.95\hsize]{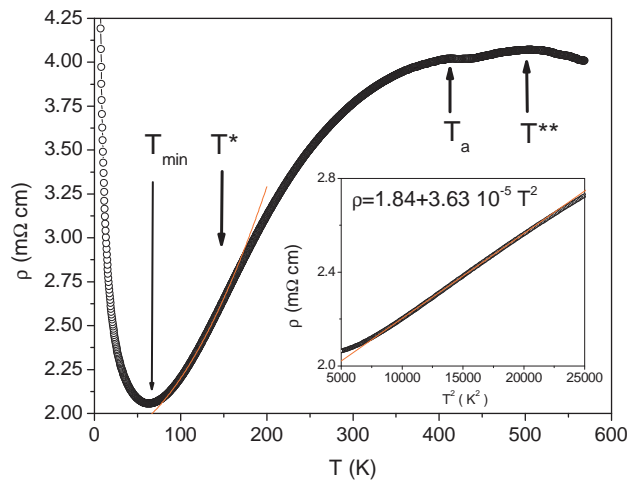}}
\caption{Temperature dependence of the in-plane single crystal resistivity. The three temperatures $T_ {min}$, $T^*$ and $T^{**}$ separate respectively 
 a low temperature insulating-like behaviour, a strongly correlated Fermi liquid, an incoherent metal and a high temperature insulator, 
while $T_a$ indicates an anomaly explained in the text. The inset displays $\rho$ {\it vs.} $T^2$ below $T^*$.}
\label{figrhoT}
\end{figure}

Specific heat measurements of sintered samples have allowed us to determine an enhanced electronic contribution 
with $\gamma=93$ $mJ/(mol.K^{2})$. A critical comparison between the Fermi liquid transport coefficient A and the Sommerfeld coefficient is made 
  with the well-known Kadowaki-Woods ratio $A/\gamma^2 $  \cite{Kado86}.
Temperature dependent Hall effect measurements have confirmed holelike charge carriers in agreement with the positive thermopower.
Both $\gamma$ and the coherence temperature $T^*$ seem to govern and thus explain the unusually high thermopower observed 
at ambient temperature in this compound.

Whereas previous transport experiments \cite{Masset00,Shikano03} was restricted to a limited temperature range, the in-plane single crystal 
 (1*0.5*0.01 mm$^3$) resistivity in Fig.~\ref{figrhoT} was measured  from 2 K to nearly 600 K using a standard four terminal method.
One can distinguish from this highly non-monotonic behaviour four transport regimes with first, 
a low temperature insulating-like behaviour characterized by a decreasing resistivity down to a 
minimum defining a transport crossover at $T_{min}\approx 63 K$. 
At higher temperatures, resistivity exhibits metallic behaviour with basically two regimes. Up to the temperature $T^* \approx 142 K$, 
one identifies a Fermi liquid regime with a resistivity varying as $\rho=\rho_0+AT^2$, as shown in the inset of Fig.~\ref{figrhoT}.
With a magnitude of $3.63$ $10^{-2}\mu \Omega cm/K^2$, A is comparable to the ones 
measured in heavy fermion compounds such $CePd_3$ for example. 
Moreover, one has to stress that the temperature $T^*$ corresponds roughly to the Mott limit $k_Fl=(h/e^2)c/\rho \sim 1$, 
with the Fermi wave vector $k_F$ and $l$ the mean free path  ($c=1.08 nm$ \cite{Masset00}). 
Therefore, $T^*$ can be interpreted as a crossover from a low temperature Fermi liquid to an incoherent metal (or "bad metal") as 
observed in the vicinity of a Mott transition \cite{Limelette03PRL}. 
Still metallic-like, resistivity in the incoherent regime increases with temperature up to $T^{**}\approx 510 K$ where it displays a 
broad maximum from which the material behaves as an insulator.

\begin{figure}[htbp]
\centerline{\includegraphics[width=0.95\hsize]{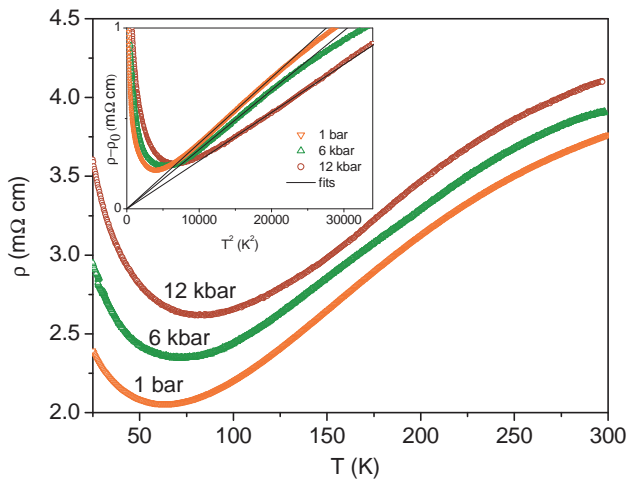}}
\caption{Temperature dependence of the in-plane resistivity under pressure, with from the bottom to the top 
1 bar, 6 kbar and 12 kbar (only three pressures are selected for clarity).
 The inset displays $\rho-\rho_0$ {\it vs.} $T^2$ with the fitted lines.}
\label{figrhoTP}
\end{figure}

Thus, one recovers the main features of strongly correlated systems near a Mott metal insulator transition observed experimentally in organic compounds 
\cite{Limelette03PRL} and predicted in the frame of the dynamical mean field theory \cite{Georges96} (DMFT).
In this theoretical context, the density of states displays  below the Fermi-liquid coherence scale $T^*$ a well-formed quasiparticle peak, 
weakly temperature dependent. In that regime the resistivity obeys a Fermi-liquid $T^2$ law. Above $T^*$, the quasi-particle resonance becomes 
strongly temperature dependent and the resistivity increases with T, reaching values exceeding the Mott limit associated to the "bad or incoherent metal" regime. 
Accordingly, the concept of well-defined and long-lived quasiparticle loses its meaning in that regime, the depletion of the density of states corresponds 
to very short quasiparticle lifetime.  
At higher temperatures $T \sim T^{**}$, the quasiparticle resonance disappears altogether, leaving a pseudogap. In this third regime, the resistivity is that 
of an insulator with a decreasing behaviour with temperature \cite{Limelette03PRL,Georges96}.

Otherwise, we note in Fig.~\ref{figrhoT} an anomaly attributed \cite{Masset00,Shikano03} to spin state transitions of the $Co^{4+}$ and $Co^{3+}$ ions  
appears at the temperature $T_a\approx 400 K$. 
The fact that $T_a$ differs from $T^{**}$ by more than 100 K suggests that these two phenomena are probably not related to one another, 
although this calls for further more specific investigations.

\begin{figure}[htbp]
\centerline{\includegraphics[width=0.95\hsize]{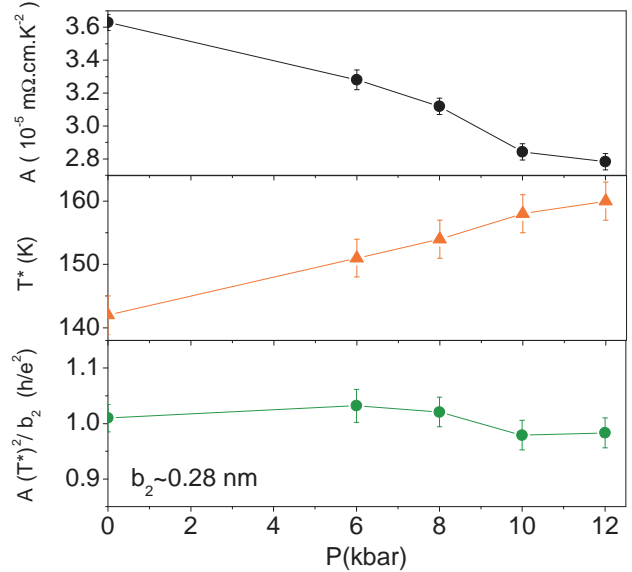}}
\caption{Pressure dependences of the Fermi liquid transport coefficient A (upper panel) and the coherence temperature $T^*$  (middle panel).
The lower panel exhibits the pressure independence of the product $A \left(T^*\right)^2/a$ in unit of $(h/e^2)$, {\it a} being an in-plane lattice parameter.}
\label{figAT}
\end{figure}

As displayed in Fig.~\ref{figrhoTP}, transport experiments under hydrostatic pressure in a clamp cell  have also been performed within the range of 6 to 12 kbar.
A striking feature of this result is that applying pressure the absolute value of the resistivity is  increasing.
Another effect of pressure is pointed out in the inset of Fig.~\ref{figrhoTP} when plotting resistivity versus $T^2$, 
with a decrease of the slope A and an increase of the temperature range, {\it i.e.} $T^*$, where $\rho \sim T^2$.
Indeed, figure \ref{figAT} confirms for all pressures investigated these two results with respectively the systematic decrease 
of the transport coefficient A while the coherence temperature $T^*$ increases with pressure.
Already experimentally observed in heavy fermions or near a Mott transition \cite{Limelette03PRL}, this quite {\it typical}  
strongly correlated Fermi liquid behaviour results basically from an increase of bandwidth due to pressure, and thus a decrease of the correlations 
which lowers the effective mass in contrast to quasiparticle coherence scale $T^*$.

An evidence for this interpretation is illustrated in Fig.~\ref{figAT} where the quantity $A(T^*)^2/a$ is plotted as a function of pressure, $a$ being 
the in-plane lattice parameter common to the two subsystems \cite{Masset00}. Indeed, this product remains constant with pressure and of order $h/e^2$ 
within the range of 1 bar to 12 kbar, confirming that both A and $T^*$ are efficient probes of electronic correlations.
Besides, we note that this result is consistent with the DMFT picture of the Fermi liquid that predicts a key role of effective mass 
$m^*$ with basically $A\sim (m^*)^2$ and $T^*\sim 1/m^*$. We stress the fact that the  pressure dependence of the lattice parameters is within 
the investigated range of pressure negligible compared to the observed variation of A and $T^*$, ensuring thus a real correlation effect.

\begin{figure}[htbp]
\centerline{\includegraphics[width=0.95\hsize]{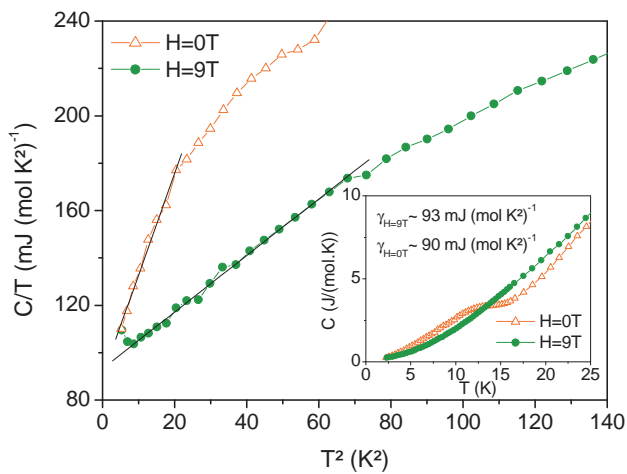}}
\caption{Temperature dependence of the specific heat for a sintered sample as $C/T$ {\it vs.} $T^2$. It's noteworthy that the 9T magnetic field 
 reveals the lattice contribution to the specific heat reducing a magnetic one \cite{Sugiyama03}, and improve the quality of the determination of $\gamma$.
The inset displays C {\it vs.} T.}
\label{figCT}
\end{figure}

In order to check the renormalization of the Fermi liquid, implying a heavy electronic mass, we have performed specific heat experiments on sintered 
sample. As observed in the inset of Fig.~\ref{figCT}, specific heat exhibits a broad maximum below 20 K consistent with the presence of magnetic fluctuations 
observed in ref.~\cite{Sugiyama03}. Therefore,  the specific heat has been measured under a 9 T magnetic field to quench magnetic fluctuations and 
reveal the $T^3$ lattice contribution on an extended temperature range. In doing this, one obtains in Fig.~\ref{figCT} a 
specific heat  behaviour as $C/T \sim \gamma + \beta T^2$ over the range 2-9 K with $\gamma \approx 93$ $mJ/(mol.K^2)$. One can emphasize 
the consistency between the latter value and the electronic contribution determined without magnetic field (Fig.~\ref{figCT}).

\begin{figure}[htbp]
\centerline{\includegraphics[width=0.95\hsize]{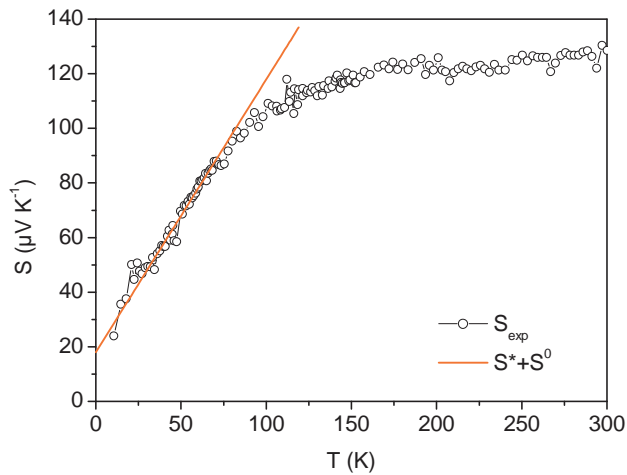}}
\caption{Temperature dependence of the single crystal thermopower S$_{exp}$ \cite{Masset00} compared to a renormalized {\it free electron} prediction 
$S^*$ involving the quasiparticle coherence scale $T^*$ (Fig. \ref{figrhoT}). Agreement is more apparent if a constant contribution 
$S^{0}$ is added to $S^*$.}
\label{figtep}
\end{figure}

We stress here on the fact that the Fermi surface seems to remain unaffected down to the lowest temperatures since there is no signature of any phase transitions 
 in both specific heat and thermopower measurements (Fig.~\ref{figtep}).
So, more than an order of magnitude higher than in conventional metals, the value of $\gamma$ attests to a strong 
renormalization of Fermi liquid parameters as the effective mass. Interestingly, the determined value is twice larger 
than in the NaCo$_2$O$_4$ compound \cite{Terasaki01}.
Consisting in a strong check of the previous conclusion, the comparison between transport and specific heat results with the so called 
Kadowaki-Woods ratio gives $A/\gamma^2 \approx  0.45$ $10^{-5} \mu \Omega cm.K^2/(mJ^2mol^{-2})$, in good agreement 
with the value found for heavy fermions  \cite{Kado86}.
Moreover, experimental value of $\gamma$ allows us to estimate the doping $\delta \sim 0.32$ using the renormalized {\it free electron} 
expression $\gamma= (\pi ^2 /2) \delta N_{av} k_B/T^*$ \cite{Ashcroft76}
, with $N_{av}$ the Avogadro number, $k_B$ the Boltzmann constant 
and $T^*$ the determined quasiparticle coherence scale instead of the standard Fermi energy.

\begin{figure}[htbp]
\centerline{\includegraphics[width=0.95\hsize]{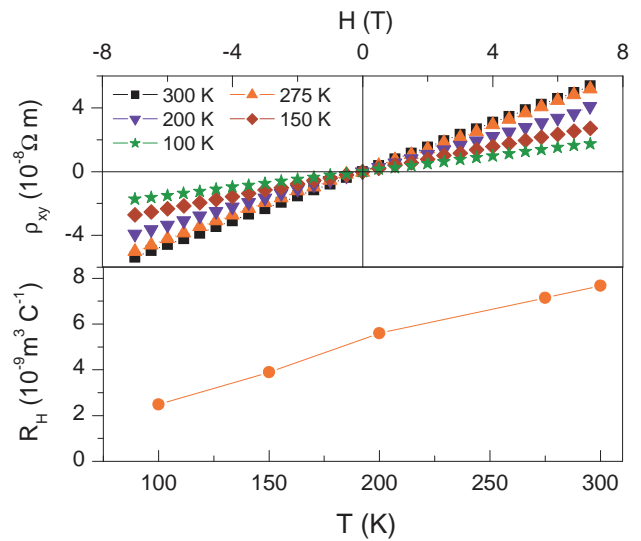}}
\caption{Magnetic field dependence of the single crystal transverse resistivity for different temperatures (upper panel) and temperature 
dependence of Hall resistance  (lower panel).}
\label{figRH}
\end{figure}

We have also performed single crystal Hall effect measurements as a function of temperature. 
We observe first in Fig.~\ref{figRH} a positive Hall resistance $R_H$ in the full temperature range investigated, implying 
thus that charge carriers are holelike. Moreover, $R_H$ exhibits globally a rather strong temperature dependence from 100 K up to 300 K, 
with an increase unusually large compared to conventional metals but consistent with the expected behaviour of strongly correlated metals  \cite{Merino00}. 

Since both $R_H$ and the thermopower S are positive, we can now discuss 
the influence of electronic correlations on thermoelectric properties in this compound.
 One may first emphasize that both S (see Fig.~\ref{figtep}) and the figure of merit $Z=S^2/( \rho \kappa)$ \cite{Koumoto02} 
 reach their maximum value at a temperature of order $T^*$ ($\kappa$ being the thermal conductivity). 
Therefore, the Fig.~\ref{figtep} clearly displays that the temperature dependence of S up to $T^*$ 
can be successfully described using a  renormalized {\it free electron} thermopower $S^*$ considering as previously, 
the determined quasiparticle coherence scale $T^*$ in eq.~(\ref{eq_S*}), with q the elementary charge.
\begin{equation}
S^*(T)= \left( \frac{\pi^2}{6} \right)  \frac{k_B T}{q T^*}
\label{eq_S*}
\end{equation}

Due to the proportionality between $S^*$ and the inverse of $T^*$, it follows from eq.~(\ref{eq_S*}) and the pressure dependence of $T^*$ 
that the low temperature slope of S should decrease with pressure. 

Furthermore, it appears on Fig.~\ref{figtep} that the agreement can be achieved up to nearly $T^*/2$ when adding to $S^*$ a low temperature constant thermopower 
$S^{0} \approx 18 \mu V/K$. 
Consequently, the thermopower S seems to result from the addition of two contributions including a constant one, namely $S^{0}$,
 in the range of temperature investigated.
While the origin of $S^{0}$ requires further experimental work, the major contribution would in the present case come from 
strongly renormalized holelike carriers giving rise to a thermopower as $S(T) \sim S^*(T)+S^{0}$ for $ T<<T^*$, 
and weakly temperature dependent for  300 K$> T >> T^*$. 
One must finally emphasize the agreement between the low temperature thermopower ($T<<T^*$) and the coefficient $\gamma$ 
with a doping $\delta \sim 0.32$ following eq.~(\ref{eq_Sgamma}) \cite{Ashcroft76}.
\begin{equation}
S(T)= \left( \frac{\gamma}{3q \delta N_{av}} \right) T+S_0
\label{eq_Sgamma}
\end{equation}

Reinforcing the interpretation of a high thermopower driven by electronic correlations, eq.~(\ref{eq_Sgamma}) confirms 
the key role of quasiparticle coherence scale $T^*$, or alternatively $\gamma$, in the electronic properties of this compound.

To conclude, both performed transport and thermodynamics measurements in the Ca$_{3}$Co$_{4}$O$_{9}$ misfit cobalt oxide 
display quite unambiguously essential features 
of strongly correlated systems, as crossovers $T^*$ and $T^{**}$, enhanced Fermi liquid parameters $A$ and $\gamma$ fulfilling Kadowaki-Woods 
ratio, while typical  pressure dependences of $A$ and $T^*$ are found. 
Furthermore, both experimental $T^*$ and $\gamma$ allow for a quantitative interpretation of the high thermopower governed 
by a strongly renormalized quasiparticles coherence scale of the order $T^*$.
In order to test the renormalized thermopower prediction, we suggest that thermopower measurements under pressure in this compound 
would consist in a key experiment for a better understanding of thermoelectricity.

\begin{acknowledgments}
We are grateful to  G. Kotliar for useful discussions.
\end{acknowledgments}

\end{document}